\documentstyle[osa,manuscript]{revtex}

\begin{document}

\begin{center}{\Large \bf
Density of States of an Electron in a Gaussian Random Potential for 
$(4-\epsilon)$-dimensional Space } \\
\vskip.25in
{I.~M.~Suslov}

{\it
P.~L.~Kapitza Institute for Physical Problems, Russian Academy of 
Sciences,\\ 117334, Moscow, Russia
} %
\end{center}

\begin{abstract}                
The density of states for the Schr{$\ddot o$}dinger equation 
with a Gaussian random potential is calculated in a space of dimension 
$d=4-\epsilon$ in the entire energy range including the vicinity of a 
mobility edge. Leading terms in  $1/\epsilon$ are taken into account for 
$N\sim 1$ ($N$ is an order of perturbation theory) while all powers of 
$1/\epsilon$ are essential for $N\gg 1$ with calculation of the 
expansion coefficients in the leading order in $N$.
\end{abstract}

\vspace{5mm}
PACS numbers 03.65.-w, 11.10.Hi, 71.23.An
\vspace{5mm}

It is a common belief \cite{1,2} that the average density of states has no 
singularity in the point of an Anderson transition in contrast to the 
conductivity and localization length. Nevertheless its calculation  
is of principal interest because all known methods  fail 
in the vicinity of the transition. Another reason is that the density of 
states and the conductivity which are determined by the average Green 
function $\langle G(x,x') \rangle$ and the correlator $\langle G^R G^A 
\rangle$ correspondingly are not completely independent quantities. In 
"parquet" approximation  difficulties in both cases are of the same character 
and are connected with  the problem of "ghost" pole \cite{3}. On the other 
hand, to satisfy the Ward identity  it is necessary to provide a strict 
correspondence of diagrams taken into account in the calculation of  self 
energy and the irreducible vertex in the Bethe--Salpeter equation \cite{4}. 
Thus no approximation for conductivity can be self-consistent before the 
corresponding approximation for density of states is formulated. This problem 
was  ignored in all existing theories \cite{3} except the symmetry approach 
suggested in Ref.~5. 

For weak disorder the mobility edge is displaced near the bare band edge
and the random potential can be regarded as Gaussian because the averaging 
is possible on scales smaller than wavelength and larger than 
the distance between scatterers (the so called  Gaussian range of spectrum 
\cite{6}). The calculation of the average Green function for  the 
Schr{$\ddot o$}dinger equation with a Gaussian random potential is reduced to 
the problem of a second-order transition with 
$n$-component order parameter 
$\vec\varphi=(\varphi_1,\varphi_2,...,\varphi_n)$ in the limit 
$n\rightarrow0$ \cite{7,8}. Then the coefficients in the Ginzburg\,--\,Landau 
Hamiltonian
$$
H\{\varphi\}=\int
d^dx\left(\frac{1}{2}c|\nabla\vec\varphi|^2+\frac{1}{2}\kappa_0^2
|\vec\varphi|^2 +\frac{1}{4}u|\vec\varphi|^4\right)
\eqno(1)$$
are related with the parameters of the disordered system by the 
relations
$$
c=1/2m, ~~\kappa_0^2=-E, ~~u=-a_0^dW^2/2,
\eqno(2)$$
where $d$ is the dimension of space, $m$ and $E$ are the particle mass and 
energy, $a_0$ is the lattice constant, and $W$ is the amplitude of the random 
potential (we set $c=1,$ $a_0=1$ in what follows). The "wrong" sign of the 
coefficient of $|\varphi|^4$ leads to unreacheability  of Wilson fixed point 
in renormgroup equations \cite{8} and to the problem of "ghost" pole in "parquet" 
approximation \cite{3}. So the possibility of $(4-\epsilon)$-expansion was  
in question for many years \cite{9} and the number of suggestions on a value 
of upper critical dimension different from four was proposed 
\cite{10,11,12,13}.  The progress was achieved in recent author's papers 
\cite{14,15,16}, where the proper treatment of factorial divergency of 
perturbation series was shown to be necessary. Here we report the results for 
$(4-\epsilon)$-dimensional case with the details of calculation to be 
published elsewhere \cite{17}.

In a four-dimensional space the structure of the perturbation series for the 
self-energy $\Sigma(p,\kappa)$ at $p=0$ has a form \cite{15} 
$$ 
\Sigma(0,\kappa)-\Sigma(0,0)=\kappa^2\sum_{N=1}^\infty u^N\sum_{K=0}^NA_N^K
\left(\ln\frac{\Lambda}{\kappa}\right)^K,
\eqno(3)
$$
where $\kappa$ is the renormalized value of $\kappa_0$ and
$\Lambda$ is the large-momentum cutoff parameter. The analogous expansion for 
$d=4-\epsilon$ has the form
$$
\kappa^2+\Sigma(0,\kappa)-\Sigma(0,0)\equiv\kappa^2Y(\kappa)=\kappa^2\sum_{N=0}^\infty
(u\Lambda^{-\epsilon})^N\sum_{K=0}^NA_N^K(\epsilon)
\left[\frac{(\Lambda/\kappa)^\epsilon-1}{\epsilon}\right]^K,
\eqno(4)
$$
where $A_N^K(\epsilon)$ are the regular functions of $\epsilon$,
$$
A_N^K(\epsilon)=\sum_{L=0}^\infty A_N^{K,L}\epsilon^L
\eqno(5)$$
and $A_0^0(\epsilon)\equiv1.$ The expansion (4) takes account of the fact 
that $Y$ is a homogenious polynomial of degree 
$N$ in  $\Lambda^{-\epsilon}$ and $\kappa^{-\epsilon}$, as follows from the 
dimensional analysis, and that the expression (4) should be reduced to (3) in 
the $\epsilon\rightarrow 0$ limit.

The quantity $Y$ satisfies the Callan\,--\,Symanzik equation \cite{18},
which follows from its relation with the vertex $\Gamma^{(1,2)}$ \cite{19}:
$$
\left(\frac{\partial}{\partial\ln\Lambda}+W(g_0,\epsilon)
\frac{\partial}{\partial g_0}+V(g_0,\epsilon)\right)Y=0,
\eqno(6)
$$
where $g_0=u\Lambda^{-\epsilon}$, $V(g_0,\epsilon)\equiv\eta_2(g_0,\epsilon),$
and the functions $\Gamma^{(1,2)}$, $W(g_0,\epsilon),$
and $\eta_2(g_0,\epsilon)$ are defined in Ref.~18. Introducing the expansions
$$
\begin{array}{c}
W(g_0,\epsilon)=\sum\limits_{M=1}^\infty
W_M(\epsilon)g_0^M=\sum\limits_{M=1}^\infty\sum\limits_{M'=0}^\infty
W_{M,M'}g_0^M\epsilon^{M'},\\
V(g_0,\epsilon)=\sum\limits_{M=1}^\infty
V_M(\epsilon)g_0^M=\sum\limits_{M=1}^\infty\sum\limits_{M'=0}^\infty
V_{M,M'}g_0^M\epsilon^{M'}
\end{array}
\eqno(7)$$
and substituting expression (4) into Eq.~(6), we obtain a system of equations 
for the functions $A_N^K(\epsilon):$
$$
(K+1)A_N^{K+1}(\epsilon)=(N-K)\epsilon
A_N^K(\epsilon)-\sum_{M=1}^{N-K}
[(N-M)W_{M+1}(\epsilon)+V_M(\epsilon)] A_{N-M}^K(\epsilon)
\eqno(8)$$
and for the coefficients $A_N^{K,L}:$
$$
(K+1)A_N^{K+1,L}=(N-K)A_N^{K,L-1}(1-\delta_{L,0})-\sum_{M=1}^{N-K}
\sum_{M'=0}^L[(N-M)W_{M+1,M'}+ V_{M,M'}]A_{N-M}^{K,L-M'}.
\eqno(9)$$

In the standard procedure of $\epsilon$-expansion \cite{7} a few first terms 
in series (5) are retained. In $M$-th order in $\epsilon$ the coefficients 
$A_N^{N-K,L}$ with  $K+L\le M-1$ are necessary for which the closed system of 
the difference equations is followed from Eq.~9. Initial conditions to this 
system and coefficients $W_{2,0},\,V_{1,0},\,\ldots$ can be derived if a few 
first orders of perturbation theory are calculated. By separating out the 
leading asymptotic term in $N$, it is easily proved by induction that
$$ 
A_N^{N-K,L}=C_{K+L}^KA_N^{N-K-L},   \eqno(10)
$$
$$
A_N^{N-K}=(-W_{2,0})^N\frac{\Gamma(N-\beta_0)}{\Gamma(N+1)\Gamma(-\beta_0)}
\frac{(-W_{3,0})^K}{(-W_{2,0})^{2K}}\frac{(N\ln N)^K}{K!},
$$
where $\beta_0=-V_{1,0}/W_{2,0}$, and the value of the first few coefficients 
in the expansion  (7) equil \cite{18}
$$ W_1(\epsilon)=-\epsilon, 
~~W_{2,0}=K_4(n+8), ~~W_{3,0}=-3K_4^2(3n+14), ~~V_{1,0}=-K_4(n+2).  
\eqno(11)
$$ 
with $K_4$ defined in Eq.~13.
For the parquet coefficients  $A_N^{N,0}$ the result (10) is exact.
The coefficients (10) do not possess the factorial growth which for $u<0$ is 
responsible for nonperturbative contribution resulting in existence 
of the fluctuational tail of the density of states \cite{14,15,16}. This 
is a reason why the usual Wilson method does not work for $u<0$. 

The approximation giving asymptotically exact results for 
small $\epsilon$ is as follows: In the lowest orders of the perturbation 
theory it is sufficient to retain in expansion (4) only leading order in 
$1/\epsilon$; for large $N$ the lowest powers of $1/\epsilon$ should be taken 
into account, since the corresponding terms grow rapidly as $N\to\infty$, but 
the leading order in $N$ is sufficient for the expansion 
coefficients in Eq.~4.  Information about the coefficients $A_N^K(\epsilon)$ 
for large $N$ can be obtained by the Lipatov method \cite{20,21}. The $N$th 
order contibution to $\Sigma(p,\kappa)$ is calculated in a close analogy with 
the $d=4$ case \cite{16} and has a form
$$
[\Sigma(p,\kappa)]_N=c_2 u^N \Gamma\left(N+b\right)
a^N\int\limits_0^\infty d\ln
R^2R^{-2}\langle\phi_c^3\rangle_{Rp}\langle\phi_c^3\rangle_{-Rp}\cdot
$$
$$
\cdot\exp\left(-Nf(\kappa R)+N\epsilon\ln R+2K_dI_4(\kappa R)\frac{1-(\Lambda
R)^{-\epsilon}}{\epsilon}\right), \eqno(12)
$$ 
where  
$$
a=-3K_4\,, \qquad b=\frac{d+2}{2}\,, \qquad c_2=c \left( 3K_4\right)^{7/2}\,,
$$
$$
f(x)=-\frac{\epsilon}{2}(C+2+\ln\pi)-3x^2\left(C+\frac{1}{2}+\ln\frac{x}{2}
\right), \qquad
\langle\phi_c\rangle_p^3=8\cdot2^{1/2}\pi^2pK_1(p),     \eqno(13)
$$
$$
I_4(x)=\bar I_4\exp(f(x)), \qquad
\bar I_4=\frac{16}{3}S_4, \qquad S_d=2\pi^{d/2}/\Gamma(d/2), \qquad
K_d=S_d(2\pi)^{-d}, 
$$
$C$ is Euler's constant,  $K_1(x)$ is a modified Bessel function, 
and  $c$ is a constant of order unity defined in Eq.~114 of Ref.~16.
Representing the result (12) in the form of expansion (4), we have
$$ 
A_N^K(\epsilon)=\tilde c_2\Gamma\left(N+b\right) a^N C_N^K
\int\limits_0^\infty d\ln R^2R^{-2}\left(\epsilon+\frac{2K_d\bar I_4}{N} 
{\rm e}^{f(R)-\epsilon \ln{R}}\right)^K\cdot
$$
$$
\cdot\exp\left(-Nf(R)+N\epsilon\ln R+2K_dI_4(R)
\frac{1-R^{-\epsilon}}{\epsilon}\right) \quad ,
\eqno(14)
$$
where $\tilde c_2=c_2\langle\phi_c^3\rangle_0^2 \approx 3.44\cdot 10^{-2}$ 
for $n=0$. 

As discussed in Ref.~16, for $d=4$ the Lipatov method reproduces 
the coefficients $A_N^K$ in (3) well only for $K\ll N$, since they 
decrease rapidly with increasing $K$ and accuracy $\sim 1/N$ of the leading 
asyptotic expression is limited. Analogous phenomenon takes place   in 
$(4-\epsilon)$ dimensions: Formula (14) is valid for all $K$ if $N\epsilon\gg 
1$ and only for $K\ll N$ if $N\epsilon \alt 1$; under these conditions the 
coefficients (14) satisfy Eq.~8, where only the term with $M=1$ is retained 
in the sum, which is possible for large values of $N$ in view of 
factorial growth of  $A_N^K(\epsilon)$. Since the system of equations (8) 
determines  $A_N^K(\epsilon)$ for $K>0$ if $A_N^0(\epsilon)$ are given,
 and the latter are well reproduced by the Lipatov method, one can determine
$A_N^K(\epsilon)$ in the region $1\ll N\alt 1/\epsilon$ for all  $K$.

Retaining in Eq.~8 only the terms with $M=1$ and $M=2$ and introducing the 
     quantity  $X_{N,M}$ by definition
$$ 
A_N^K(\epsilon) = [-W_2(\epsilon)]^K \frac{\Gamma(N- 
\beta(\epsilon))}{\Gamma(K+1) \Gamma(N-K-\beta(\epsilon))} 
A_{N-K}^0(\epsilon)  
X_{N,N-K}\eqno(15) 
$$ 
where $\beta(\epsilon) = -{V_1(\epsilon)}/{W_2(\epsilon)} $, we have an 
equation
$$ 
X_{N,M}=\left( \hat l_M +\hat \delta_M \right) X_{N,M+1} \quad,\eqno(16)
$$
 where 
$$
\hat l_M \equiv h_{M} + {\rm e}^{-i\hat p} \quad , \qquad       
\hat \delta_M
 \equiv \frac{f_{M}}{N} {\rm e}^{-2i \hat p} \quad \eqno(17)
$$
$$
h_{M} = -\frac{\epsilon}{W_2(\epsilon)} 
\frac{A_{M+1}^0(\epsilon)}{A_M^0(\epsilon)} 
\frac{M+1}{M-\beta(\epsilon)}\quad ,\qquad 
f_{M}= \frac{W_3(\epsilon)}{W_2(\epsilon)}  
\frac{A_{M-1}^0(\epsilon)}{A_M^0(\epsilon)}(M-1- \beta(\epsilon))
\eqno(18) 
$$ 
and ${\rm e}^{-i\hat p}$ is the displacement operator on the distance $-1$ 
affecting the both arguments. The equation (16) can be formally solved by 
straigtforward  iterations taking into account  the boundary condition 
$X_{N,N}=1$. The result can be expanded in powers of  $\hat \delta_M$, 
terms containing a few number of operators $\hat \delta_M$ being essential.
These terms can be explicitly calculated by induction.
For $M\sim 1$ we have a result
$$
A_{N}^{N-M}(\epsilon) 
 = \frac{1}{M!} \epsilon^{N-M} \tilde c_2 \Gamma(N-\beta(\epsilon)) a^N
 \left( t/2\pi \right)^{1/2}
{\rm e}^{ f_{\infty} (Nt\ln N -1) +1/t} \cdot
$$
$$ \qquad \qquad
\cdot  \int_0^{\infty} dx {\rm e}^{ - \frac{t}{2} (N-1/t-x)^2 }
  x^{M+b+\beta-f_{\infty}Nt} J(x)
\quad , \eqno(19)
$$
where 
$$
J(N) = \int\limits_0^\infty d\ln R^2R^{-2}
\exp\left(-Nf(R)+N\epsilon\ln R+2K_dI_4(R)
\frac{1-R^{-\epsilon}}{\epsilon}\right) \quad ,
\eqno(20)
$$
and
$$
t = -\frac{\epsilon a} {W_2(\epsilon)}
\stackrel{\epsilon \to 0} {\longrightarrow} \frac{3\epsilon}{n+8}
   \quad , \qquad
f_{\infty} = \frac{W_3(\epsilon)}{aW_2(\epsilon)} 
\stackrel{\epsilon \to 0} {\longrightarrow} \frac{3n+14}{n+8}
  \eqno(21)
$$
which is valid in the region  $Nt>1$ or $1-Nt \ll \epsilon^{1/2}$.
For $N\epsilon \ll 1$ and $N-K\ll \ln N$ we have a result of type (5) with 
$A_N^{K,L}$ given by (10). Other regions of parameters can be  also
investigated but they do not make essential contributions to the sum 
of the perturbation series.

Two contributions are important in the sum 
of Eq.~4:  (a) nonperturbative 
contribution 
$$
[\Sigma(0,\kappa)]_{nonpert} \equiv i {\mit \Gamma}_0(\kappa^2)
=i\pi\tilde c_2 \kappa^2 \left(\kappa
^{\epsilon}/au \right)^b {\rm e}^{-\kappa^{\epsilon}/au}
F \left( \kappa^{\epsilon}/au  \right)
\eqno(22)
$$  
(the limit $\Lambda\to\infty$ is taken)
arising from the region of large $N$ and obtained from Eq.~4 by summation 
over $N$ from an arbitrary finite $N_0$ to infinity, if the coefficients 
$A_N^N(\epsilon)$ given by (19) are written in the form
$$
A_{N}^{N}(\epsilon) =  \tilde c_2 \Gamma(N+b) {\epsilon}^N
a^N F(N) \eqno(23)
$$
(the formula (46) of Ref.~16 is used); (b) the quasiparquet 
contribution arising from the terms with coefficients $A_N^{N-K,L}$ with 
$K\sim L \sim 1$ given by (10)
$$
[Y(\kappa)]_{quasiparq}=\left[\Delta+\frac{W_{3,0}}{W_{2,0}}u
\kappa^{-\epsilon}\ln\Delta\right]^{\beta_0},
~~\Delta\equiv1+W_{2,0}u\frac{\kappa^{-\epsilon}-\Lambda^{-\epsilon}}{\epsilon}.
\eqno(24)
$$
To logarithmic accuracy, the quantity $\Delta$ in the logarithm
can be replaced by its minimum value  $\bar\Delta\sim\epsilon\ln\epsilon$ 
(determined by Eqs.~ (26)\,--\,(30) presented below), and in the limit  
$\Lambda\rightarrow\infty$ the result (24) can be written in the form
$$
[Y(\kappa)_{quasiparq}=[1+W_{2,0}\tilde u\kappa^{-\epsilon}/\epsilon]^{\beta_0},
~~\tilde u\equiv u\left[1+\frac{W_{3,0}}{W_{2,0}^2}\epsilon\ln\bar\Delta\right],
\eqno(25)
$$
which differs from the parquet result \cite{19} only in that $u$ is 
replaced by $\tilde u$. It can be proved \cite{16} that such replacement occurs in 
all parquet formulas employed in calculating the density of states \cite{15}.

The rest of the calculations are similar to those described in Ref.~15.
The damping  ${\mit \Gamma}$, the renormalized energy $E$, and the density of 
states $\nu$ are determined in parametric form as functions of the bare 
energy $E_B$ by the equations 
$$
{\mit \Gamma}={\mit \Gamma}_c\left(1+\frac{\epsilon x}{2}\right)^{2/\epsilon}\sin\varphi,
~~E=-{\mit \Gamma}_c\left(1+\frac{\epsilon x}{2}\right)^{2/\epsilon}\cos\varphi,
\eqno(26)$$
$$
-E_B+E_c={\mit \Gamma}_c\left(\frac{\epsilon
x}{2}\right)^{1/4}\left(1+\frac{\epsilon
x}{2}\right)^{2/\epsilon-1/4}
\left(\cos\left(\varphi+\frac{\varphi}{4x}\right)-
{\rm tg}\frac{\varphi(1+2\epsilon
x)}{3}\sin\left(\varphi+\frac{\varphi}{4x}\right)\right),
\eqno(27)$$
$$
\nu=\frac{{\mit \Gamma}_c}{4\pi|\tilde u|}\left(1+\frac{\epsilon
x}{2}\right)^{2/\epsilon}\left(\left(1+\frac{2}{\epsilon
x}\right)^{-1/4}\sin\left(\varphi+\frac{\varphi}{4x}\right)\left[1-\frac{R_0^2}{2(1+\epsilon
x/2)}\right]-\right.
$$
$$
\left.-\left(1+\frac{2}{\epsilon x}\right)^{-3/4}\sin\left(\varphi+\frac{3\varphi}{4x}\right)\right),
\eqno(28)$$
$$
{\mit \Gamma}_c=\left(\frac{8K_4|\tilde u|}{\epsilon}\right)^{2/\epsilon},
~~E_c\simeq 2u\int\frac{d^dk}{(2\pi)^d}\frac{1}{k^2}\quad, \qquad
R_0\approx \left( \frac{\epsilon}{3\ln(1/\epsilon)}\right)^{1/2}
\eqno(29)$$
where $x(\varphi)$ is a single-valued function in the interval 
$0<\varphi<\pi$, similar to the function shown in Fig.~2 of Ref.~15 and 
determined by equation
$$
\sin\left(\varphi+\frac{\varphi}{4x}\right)=\frac{e^{-4x/3}}{x^{1/4}}
I(x) \cos\frac{\varphi(1+2\epsilon x)}{3} \eqno(30)
$$
where
$$
\! I(x)= \tilde c_2 \left( \frac{3}{4} \right)^{1/4}
\left( \frac{\pi t}{2} \right)^{1/2}
{\rm e}^{-f_{\infty}+f_{\infty} (1+\epsilon x/2) 
\ln{ [\bar \Delta (1+\epsilon x/2)/t]}}  
  \int_0^{\infty} \! dz 
{\rm e}^{-\frac{t}{2}  \left(\epsilon x/2t-z \right)^2 }
z^{b+\beta-f_{\infty}(1+\epsilon x/2)} J(z)  
   \eqno(31)
$$
The expressions (26--30) are simplified drastically in two 
overlapping regions. For large $|E|$, when $x\gg \ln{(1/\epsilon)}$, the 
right-hand side of (29) is small and the quantity  $\varphi$ is close to
0 or $\pi$, so for  ${\mit \Gamma}(E)$ and $\nu(E)$ we have asymptotic 
expressions 
$$
{\mit \Gamma} (E) = \left \{ \begin{array}{cc}
{\displaystyle\frac{\pi}{8} }
\epsilon E \left[ (E/{\mit \Gamma}_c)^{\epsilon/2}-1 \right]^{-1}
\quad ,
 & E \gg  {\mit \Gamma} \\
{\mit \Gamma}_0(E) \left[ 1- (|E|/{\mit \Gamma}_c)^{-\epsilon/2} \right]^{-1/4} \quad ,
& -E \gg  {\mit \Gamma} 
\end{array} \right .    \quad , \eqno(32)
$$
$$
\nu (E) = \left \{ \begin{array}{cc}
{\displaystyle\frac{1}{2}} K_4 E^{(d-2)/2} \left[ 
1-\left(
{\displaystyle\frac{E}{{\mit \Gamma}_c}} \right)^{-\epsilon/2} 
\right]^{-1/4} \,\, ,
 & E \gg  {\mit \Gamma} \\  {    }& {       }\\
{\displaystyle\frac{{\mit \Gamma}_0(E)}{4\pi |\tilde u|}} \left( 1-
{\displaystyle\frac{R_0^\epsilon}{2}  }
\left( {\displaystyle\frac{|E|}{{\mit \Gamma}_c}} \right)^{-\epsilon/2}
-\left[ 1-\left( {\displaystyle\frac{|E|}{{\mit \Gamma}_c}}
\right)^{-\epsilon/2} \right]^{1/2}
                                                   \right) \,\, ,
& -E \gg  {\mit \Gamma} 
\end{array} \right . \quad , \eqno(33)
$$
(${\mit \Gamma}_0(E)\equiv {\mit \Gamma}_0(|\kappa|^2)$) which give 
an illusion of ghost 
pole \cite{3}.  For small $|E|$, when $x\alt {\epsilon}^{-1/2}$, Eq.~29 
takes the form 
$$ 
\mbox{sin} (\varphi + \varphi/4x) = I(0) \mbox{cos} 
(\varphi/3) \frac{e^{-4x/3}}{x^{1/4}} \quad ,\qquad I(0)\sim \epsilon^{-7/12} 
\left( \ln \frac{1}{\epsilon} \right)^{17/12} \quad , 
\eqno(34) 
$$ 
and the ghost pole is shifted from the real axis into the complex plane on 
the distance  $\sim \epsilon {\mbox{ln} (1/\epsilon)}$.

For large positive $E$  Eq.~32  gives the inverse relaxation time 
appearing in  the kinetic equation  while for large negative $E$
the damping ${\mit \Gamma}$  becomes purely nonperturbative. The function 
$\nu(E)$ for large positive $E$ goes over to the density of states 
of an ideal system, and for large negative $E$ the following result is 
obtained for the fluctuational tail
$$ 
\nu(E) =\frac{K_4}{\pi} {\mit \Gamma}_0(E) |E|^{-\epsilon/2} \ln\frac{1}{R_0} 
=\tilde c_2 K_4\left(\frac{2\pi}{3}\ln\frac{1}{R_0}\right)^{1/2} 
R_0^{-3}|E|^{(d-2)/2} \left[\frac{\bar I_4 |E|^{\epsilon/2}}{4|u|}\right] 
^{(d+1)/2} \cdot
$$
$$
\cdot \exp\left(\frac{2K_dI_4(R_0)}{\epsilon}-\frac{I_4(R_0)|E|^{\epsilon/2}}
{4|u|R_0^\epsilon}\right),
\eqno(35)
$$
the energy dependence of which is identical to that obtained in Ref.~22 and 
23 and corresponds to Lifshitz's law \cite{6}; the divergence in the limit 
$\epsilon\rightarrow0$ is removed for a finite cutoff parameter $\Lambda$.   

It is interesting, that for $\epsilon x\ll1$ formulas
(26--30) have the same functional form as in the $d=4$ case \cite{16}, i.\,e. 
the behavior of all physical quantities in the vicinity of a mobility edge is 
effectively four-dimensional.  As in Refs.~15 and 16, the phase transition 
point shifts into the complex plain, and the density of states has no 
singularities for real $E$ in accordance with widely accepted but not proved 
ideas.

I am grateful to the participants of the seminars at the Institute for 
Physical Problems and the Physics Institute of the Academy of Sciences
for their interest in this work.

This work was performed with the financial support of the International 
Science Foundation and the Russian government (grants MOH000 and MOH300)
and the Russian Fund for Fundamental Research (grant 96-02-19527).

\end{document}